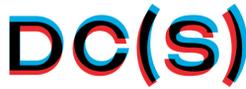

# A Design Experience for Interactive Narrative Based on The User Behavior

**Yuan Yao[a][b], Haipeng Mi*[a][b],**

[a] Department of Information Art and Design, Tsinghua University
[b] The Future Laboratory, Tsinghua University
* haipeng.mi@acm.org

**Abstract** | Research on interactive narrative experiences in physical spaces is becoming more popular, growing into an established new media art format with the development of technology and evolution of audience aesthetics. However, the methods of designing interactive narratives are still similar to the basic video narratology of traditional designers, directors, and producers. This paper provides a design method based on the user's physical behavior and proposes an art installation by this method, where the aim of the installation is to transmit a more vivid story to users, presenting a new research inspiration of interactive narratology for designers and researchers.

# 1. Introduction

The intervention of technology makes the story more vivid and even developed a new narrative method, named interactive narrative. However, research on interactive narrative is mostly theoretical such as (Gerrig, R, 2018), (Gervás, P, 2009), (Weyhrauch, P, 1997), and (Louchart, S. et al, 2004). even in 1970, Roger Schank had already proposed the prototype of interactive storytelling (García, R et al, 2014). Although recently, there has also emerged design practice, most of which are interactive computer programs focused on story plots, for instance (Ryan, M. L, 2008), (Delmas, G, 2007) and (Riedl, M. O. et al, 2013). Taking into consideration of user behavior interaction in physical space, we can also see some cases of interactive installation design, most of which are advertisements to attract users to participate with interaction and incentive mechanisms (Ojala, T. et al, 2012), (Zhang, Y.et al, 2018) and (Müller, J. et al, 2012). In recent years, there have been a lot of interactive narrative cases combined with VR, which is a good start to focus on user behavior, and makes possible the user's immersive experience[1] such as (Bates, J, 1992). However, VR is not a real physical experience but a way of interaction. The development of technology may give us access to multimodal sensors and experience more diversified stories. Because the essence of interactive narrative is both the user's immersion and fascinating stories. *Sleep No More*[2], an interactive stage play, sets a good example of excellent interactive experience.

This paper elaborates a design case of interactive narrative combining user behavior in physical space. In the course of our research, we attempt to find the mapping relation between user's natural behavior and narrative camera language, and to explore the possibility to apply it to an interactive narrative on the large display.

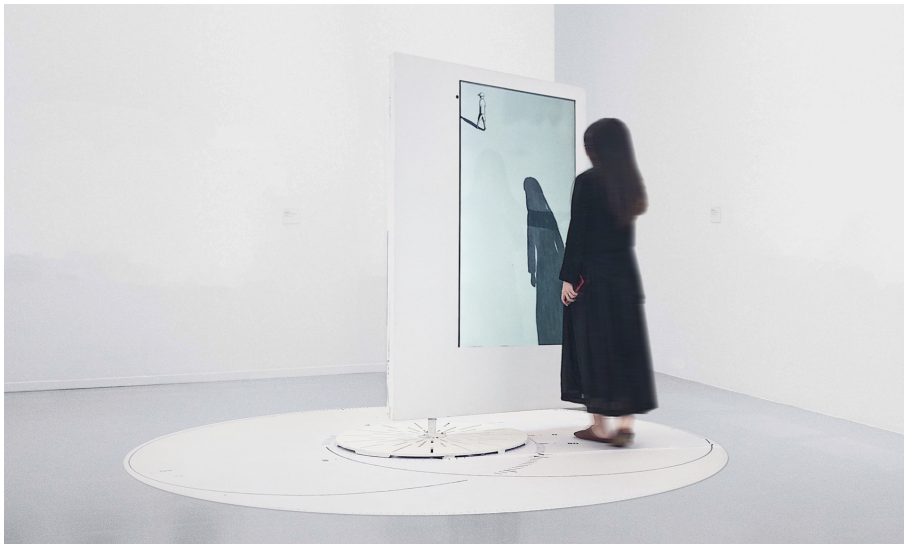

*Figure 1.*      *Interactive Installation — Still Walking*

---

[1] The VR game, Tree, https://www.media.mit.edu/posts/tree-treesense/
[2] The interaction drama, Sleep no more. https://mckittrickhotel.com/

An interactive installation, *Still Walking*, is designed to explore the spatial interaction between the user and the story. The installation's narrative theme is based on temporal change, while the concept of "shadow" acts as the narrative glue to connect the story with the user. As a user approaches the installation screen, the shadow of a woman appears on an endless path. If the user revolves the screen and moves forward, the shadow moves along, the scene changing bit by bit from a child to an older woman. In this process, branching stories at different life stages—determined by the age—of the older woman are triggered by the different positions of the screen. We are taking advantage of this installation to present users with a novel and thought-provoking interactive experience. In this paper, I would dig into the process of designing this installation as well as our various thoughts and takeaways on interactive narrative design.

## 2. Narrative Design

### 2.1 Interactive narrative structures

In the traditional film narratology, the narrative structure is linear. Even it has enriched plots. However, the interactive narrative presents a broader range of plot types. It is typically composed of a structure where has multiple starting points that may lead to various outcomes, as illustrated in tree-structure, network structure, and etc. (Figure 2)

In *Still Walking*, the storyline is organized in the circular narrative structure. There are different branching plots in the entire story, and the interaction leads to different endings. The narrative structure is a similar circular network. (Figure 3)

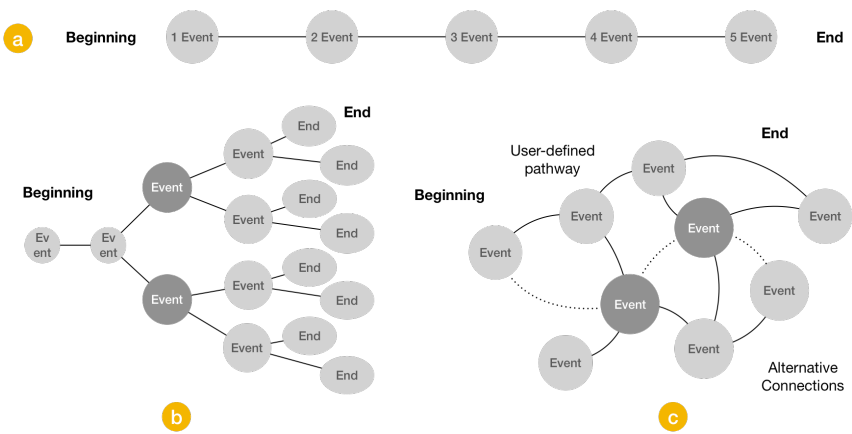

Figure 2.     (a)The traditional narrative structure; Interactive narrative structure, tree (b) and network(c);

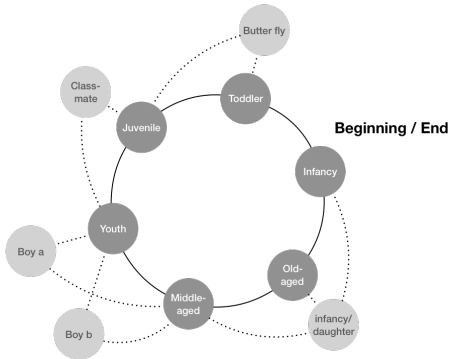 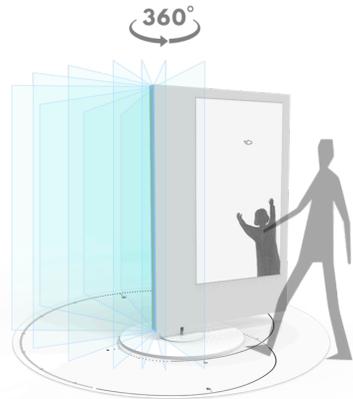

*Figure 3. The narrative structure of Still Walking*     *Figure 4.    The concept of interaction*

## 2.2 Interaction Design

The core purpose of narration is to establish an emotional connection after the audience has watched and understood the story (Bordwell, D, 2012). This can be realized, when there are clues weaved in the story, guiding the audience to piece together and establish emotional resonance with the whole story. In this design, we try to give back the interaction rights to the audience, make them users. According to the interactive plots and clue to guide the users to explore the story and build emotional connections.

The story of *Still Walking* is a life circle, and the users also need to come back to beginning point in the end. Therefore, we designed a circle road, utilized rotation and pause as the gates of the other plots. Additionally, the rotation of the screen brings out the shadow as a story clue moving along with the user, and guide the users by different states. (Figure 4)

## 2.3 Story script

The story concerns a woman's lifetime, which is divided into four sections. One rotation of the installation means the story completes for one time, the shadow changing form a girl to an old woman (Figure 5), which takes about 40 seconds. As the installation rotates, the user goes through the four periods in her life, namely infancy, juvenile, youth, middle-aged and old-aged life, each scene lasting 10 to 15 seconds. In other words, the interaction between the installation and the user lasts about 40 to 100 seconds. The whole story, with the help of mimicry and fixed camera angle, is unveiling a woman's growth, symbolized by the changes of the shadow. (Figure 6)

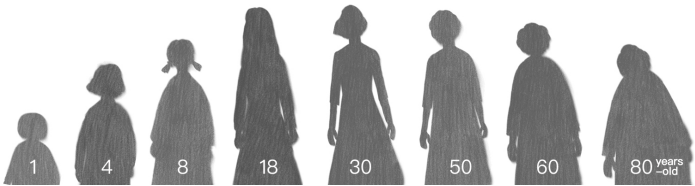

*Figure 5.*    Images of women at different ages

### Infancy: a carefree catch

The 4-year-old protagonist, no longer a toddling infant, is jumping joyfully. Suddenly, a butterfly appears ahead of the road.

- Keep walking: The girl is running after the butterfly and reaching her hands to catch it.
- Stop walking: As she stops, the butterfly flies away.

### Juvenile: a friendly companion

The eight-year-old protagonist sees a girl in her class with a backpack.

- Keep walking: She waves her hands and catches up with her classmate. Then walking together, until her classmate's shadow fades away and finally disappears.
- Stop walking: If she stops, she can never catch up with her classmate.

### Youth: a difficult choice

The eighteen-year-old protagonist finds the boy she likes. One of her hands is reaching to him but pauses halfway. She is about to chase after him.

- Keep walking: No matter how fast she runs, she can't catch up with him.
- Stop walking: Another boy will appear behind her. Accompanying her for a while, the shadow of the boy fades away and vanishes, leaving the girl to go forward herself.

### Middle-aged and old-aged life: an affectionate change

The forty-year-old protagonist sees her little daughter walking in front of her. She is waving her hands to get her daughter's attention and wants to catch up with her toddling daughter. As the screen rotates and the user keeps walking, she realizes her daughter's change is much slower than hers, but then she ages quickly from 50 years old to 80 years old.

- Keep walking: As the screen keeping rotating, her daughter is getting close to the camera, while the shadow of the protagonist is backing out and fading away. Push the screen to the original position, and the protagonist becomes her daughter and starts a new circle of life over again.

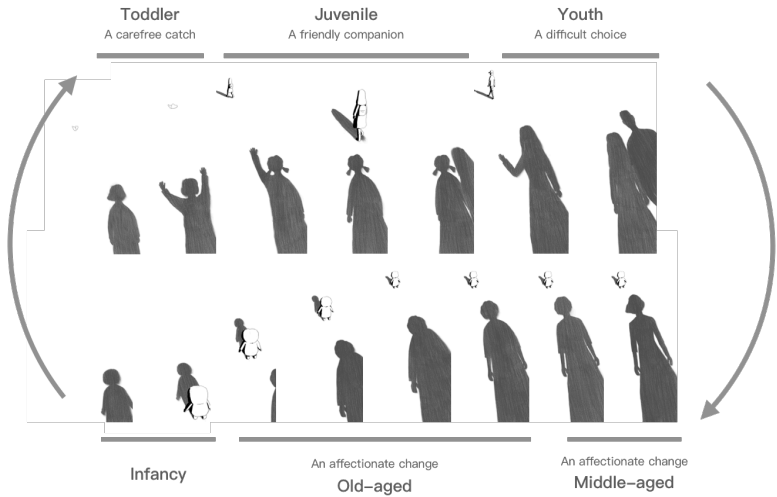

*Figure 6.     Overview of the story*

## 3. Mapping Behavior and Camera Language

### 3.1 User behavior in interactive narrative

Created for the 2015 Melbourne Writers Festival and by J. Walter Thompson, Twists & Turns[1] is not only an app but also a case of large-scale, outdoor interactive narrative. This crafty design uses the location, direction, and movement of the user in a large space to deliver a story in which the user can choose the branching plots. To be specific, each street and every corner in Melbourne become twists and turns in the story. This case is an indirect proof that the user's natural behavior serves not only as an inspiration in the design process but also as a chance to develop a new interactive form, which is easy for the user to interact with and for the designer to create. The proxemic in (Ballendat, T. et al, 2010), (Hall, E. T, 1966) and (Vogel, D. et al, 2004) offers us a perspective that the user's natural behavior has both implicit and natural interaction with the installation in physical space. Based on this which many design forms of natural interaction are created for the users. Throughout our research, we take five factors into consideration to explain user behavior: distance, identity, location, direction, and movement. (Figure 7)

---

1 Twists & Turns, https://andyhincey.com/portfolio/melbourne-writers-festival-twists-turns/

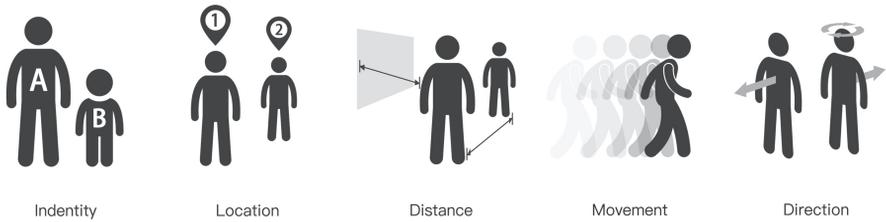

*Figure 7.    The 5 factors of the user behavior*

In this research, we would use the camera language in filmology as the main content for mapping relationships. The camera language provides an opportunity for user behavior, enabling the user to participate in the plot. In traditional film narrative, the camera language includes the camera shot, angle, camera movement, and montage. The only way for the user to understand the camera language is to "watch". However, in this paper, we try to establish a mapping relation between the five aforementioned factors of user behavior and the narrative camera language, which provides the user with a vivid way to understand the story.

## 3.2 Distance, position and direction of the user VS. Camera shot and angle of the camera

The camera shots of the camera language, in a broad sense, include long shots, full shots, medium shots, and close-ups. Taking narrative patterns into consideration, there is a different system of shots, such as the detail shot, narrative shot, and environment shot. The wider the camera shot, the less the story tells; the closer the camera shot, the more the story expresses. There are also have different camera angle types, such as eye-level shot, low-angle shot, high-angle shot as well as frontal, side, and canted angle. Different combinations of camera shot and angles in traditional film narratives can present different levels of story content for the audience.

At the same time, research on large-display interaction indicates that the user's intention will be stronger as the distance is closer, that mentioned by (Ballendat, T. et al, 2010), (Michelis, D, 2011) and (Vogel, D. et al, 2004). Therefore, we combine the camera shot and angle of narrative language with the distance, position, and direction of the user in the narrative language. The change of user behavior may trigger the transform of camera shot and angle, reinforcing the user's experience.

We have two hypotheses. 1)The mapping between the user's distance and the camera shot: the farther the user keeps from the screen, the wider the camera shot changes, and it means less content in the story. On the contrary, the closer the distance is, the closer the camera shot changes, and it means more content in the story. 2) The position and direction of the user will determine the viewing angle, which has a mapping with the shooting angle. The change of the user's position results in the change of point of view in the story, which can improve the user's experience of story understanding.

## 3.3 Movement of the user VS. Camera movement and montage of the camera language

Besides the basic cinematography movements, namely zoom, pan, follow, tilt, dolly, truck, pedestal and rack focus, there are two basic forms of motion—mimicry, and spontaneity (table 1). Mimicry indicates the movement of a camera that follows the action of the character or other objects in the story, which makes it easier for users to participate as a character. Spontaneity is more about watching the development of the plot from a third-person point of view.

*Table 1. Forms of camera motion*

| Forms of motion | Moving subject in the film | Camera movement |
|---|---|---|
| Mimicry | （1）Object | Stationary |
| | （2）Background | Synchronous |
| | （3）Line of sight | Mimic |
| Spontaneity | （4）Camera | Spontaneous |

The camera movements can be combined with the user's physical movements. In *Still Walking*, the user's movement, or rather rotation, is designed to mapping with the "follow" of camera movement. The camera shooting the story moves along as the user continuously pushes the screen forward. Once the user stops, the camera stops as well. What's more, the shooting movement also changes with the speed at which the user pushes the screen.

Also, montage is a major feature of film narrative, using various combinations of images to stimulate emotions and lead the plot to a climax. In the interactive narrative, we attempt to apply the montage to piece together separate sections in different scenes into a continuous whole, for example, similar movement of users and camera shots.

## 3.4 Identity of the use VS Clue of the story

*Table 2. The mapping relation between user behavior and camera language*

| User behavior | Mapped as | Camera language |
|---|---|---|
| Identify | Clue | Clue |
| Location/Direction/Distance | Point of view | Angle and camera shot |
| movement | Motion (mimicry, spontaneity） | Camera movement |
| Above all | Story flow | Montage |

In the story, a user makes use of the clues to thread the plots, and in the same way, the interactive installation in the physical space can also use the clues to guide the user to explore the branching plots. And the clues can mapping with the user identity. While in *Still Walking*, we take the shadow of the user as its clue, and the user can quickly get themselves involved in the story as the protagonist. (table 2)

# 4. Implementation

Compared with the traditional narrative, the interactive narrative depends more on technology to convey information that facilitates interaction between the user and story. The interactive technology supports tasks such as user behavior perception, data analysis, and information feedback. Interactive technology plays a supporting role in installation design, where we should focus on the theme of the work and procure suitable technology to present information.

## 4.1 Mock-Up

The technical design has a complicated process, and we have to predict hidden problems that may appear at different stages. Thus, it's necessary to have a mock-up before the formal implementation. We conducted a lot of tests during this phase, such as screen size, animation type, sensors, and main supporting structures.

In this phase, the main content was designing the animation composition. We stuck a 2-squares meter paper on the wall for testing. The first was to measure the distance between the user and the screen. In interaction, the user needs to push the installation to rotate, the distance between them is quite close. If the scene size is too large, it will be distorted and cause the vision burden to users. Therefore, scene composition needs to conform to the comfortable perspective, contain the size and position of shadow, and other characters.

The primary hardware is display device. We compared the effect of projection and LED screen. The projection can be resized, but it needs more space. The advantage of the LED screen is high-definition, but it cannot be resized and much heavier. However, we choose a large-size LED screen in the end, that means we have to consider the stability of the structure during the implementation.

## 4.2 Interactive technique

The core technique consists of two parts: the hardware and the software. The hardware here is a 65-inch TV screen rotating 360-degrees around a vertical axis and it is the main body of the installation. We have paid great attention to the safety issue and its power supply. For safety concerns, we have added extra weight on both the shaft and base to prevent the screen from falling. It's a steel plate with 1 m diameter and 3 cm thickness. For the supply of electricity, an industrial-grade slip ring allows the transmission of power. And we designed a special structure to break down the tension generated by rotation to ensure the safety of the slip ring. (Figure 8)

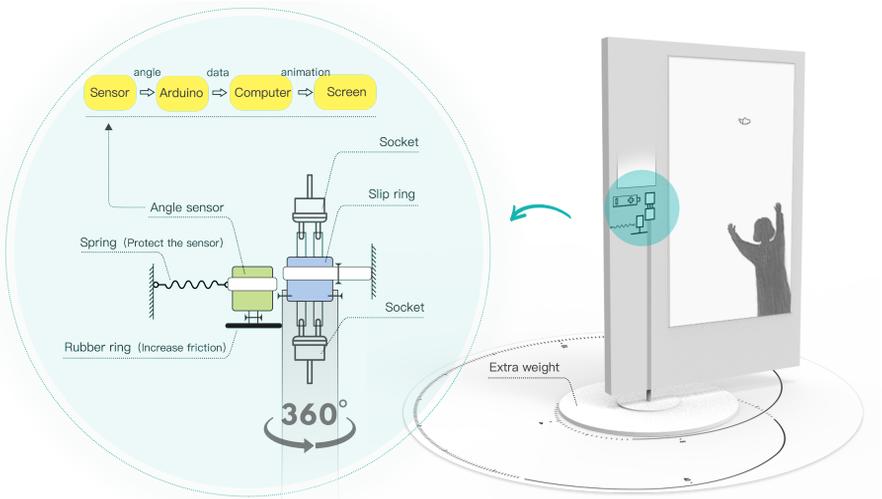

*Figure 8.    Technological principles in Still Walking*

In software design, there are two pieces involved: Angle Sensor is used to register the angle data of rotation and transfer it to computer by Arduino. After that, the Unity3D would render the sequences of animation frames that according to our preset program. (table 3)

*Table 3. The numbers of animation frames in each plot.*

| Age | 1 years old | 4 years old | 8 years old | 18 years old | 30 years old | 90 years old |
|---|---|---|---|---|---|---|
| Age Changes | 30 | 35 | 34 | 30 | 66 | |
| Walk | 0 | 12 | 35 | 20 | 65 | |
| Pause | 4 | 4 | 4 | 4 | 4 | 4 |
| Wave Hands | 0 | 12 | 19 | 20 | 24 | 0 |
| Other Characters | 0 | Butterfly 31 | Classmates 10-35 | Boy_a 20 Boy_b 22 | Baby 4-60 | |

## 4.3 Installation

The installation has 1.8m high, 1.5m wide, and covers an area of 3m by 3m. And add ground stickers with age symbols. All surfaces of the installation are painted white to increase interactive immersion. It has a circle motion, the user can stand around and join it anytime and any angle. This display mechanism is quite light and easy to push, however, considering the safety of many young users, we have specially posted a warning banner.

In the interactive narrative, the design process has multiple feedback. Besides the visual feedback, it may get other sensory feedback concerning auditory sense, tactile sense, olfactory sense, and etc. All the feedbacks are using to enriches the users' experience and helps them to enjoy the story.

## 5. Observation and improvement

During the exhibition, more than thousands of people had experienced the interaction. Even there have audiences posted the video on social networks and watched by millions of people. Most of them would praise the interaction and moved by the story. However, there still have users doubts about the interactive method of the installation if there were no accompanying instructions. It is not easy for the users to figure out that they should interact with the installation through "push". People at different ages respond to the installation in different ways: children from 3 to 10 years old prefer active engagement, but most of them are attracted by the revolving screen and fail to explore it further. Young people from 15 to 25 years old make up their minds on whether or not to take the next step after watching how others may respond first. Middle-aged people over 40 prefer watching to joining. Based on user behavior in the first round, we put up signs "Push slowly" on the frame of the installation, which turns out to be a good reminder for young people and middle-aged people. For the aesthetic, the notice should be designed as part of the installation or put on the ground, rather than putting it up on the frame.

As for the story content design, some users failed to find out the branching plots. Even though they were informed of the interactive method in advance, some users still could not find the proper angle to stop the screen. Learning from this, we need a more attainable angle to trigger the interaction, and users should also have proper guidance on how to interact with the story. Despite all the defects above, the content and form of our story have been recognized by the users. The theme of time and lifespan was clear enough to be consistently understood by the users.

## 6. Conclusion and Future Work

We have discussed the feasibility of using user behavior as an interactive trigger on the large display. And we prove it can not only help the users to understand the story, but also make the installation more attractive and immersive.

In this case, we discuss the situation of one single person as part of the interactive narrative. However, it can also involve other users. The interaction among them and their user behavior will inevitably change the interactive storytelling.

In addition, we only discuss the interaction in a physical space. The future interactive installation may relate different spaces together, and the users' behavior in different spaces may trigger the interaction mechanism so that they can participate in the story together.

Finally, the interactive narrative method based on user behavior proposed in this paper is only an experimental exploration. It is just the tip of the iceberg in terms of studies on the new media and interactive narrative. We hope this case serves as a good start for further exploration of the relationship between narrative and user behavior.

**About the Authors:**

**Yuan Yao** is a Ph.D. candidate at the Department of Information Art and Design, Tsinghua University and The Future Laboratory, Tsinghua University. Her research is focused on the future of human's interactive experiences with environments. In particular, she is interested in the interactive narrative that engages in New Media Art and Public space.

**Haipeng Mi** is an Associate Professor at the Department of Information Art and Design, Tsinghua University. He is also a visiting researcher in the Future Laboratory of Tsinghua University. His research interests range from user interface design, human-robot interaction to interactive art. He has over 30 peer-reviewed publications in the field of HCI.